\documentclass[12pt]{article}

\usepackage{epsf}
\usepackage{epsfig}
\usepackage{cite}
\usepackage{amsmath}
\usepackage{graphics}

\begin{document}

\setlength{\textheight}{21.5cm}
\setlength{\oddsidemargin}{0.cm}
\setlength{\evensidemargin}{0.cm}
\setlength{\topmargin}{0.cm}
\setlength{\footskip}{1cm}
\setlength{\arraycolsep}{2pt}

\renewcommand{\thefootnote}{\#\arabic{footnote}}
\setcounter{footnote}{0}

\newcommand{\gtrsim}{ \mathop{}_{\textstyle \sim}^{\textstyle >} }
\newcommand{\lesssim}{ \mathop{}_{\textstyle \sim}^{\textstyle <} }
\newcommand{\rem}[1]{{\bf #1}}
\renewcommand{\thefootnote}{\fnsymbol{footnote}}
\setcounter{footnote}{0}
\def\thefootnote{\fnsymbol{footnote}}

\begin{titlepage}

\hfill {\tt December 2018}\\

\begin{center}
{\Large \bf Possible Bilepton Resonances \\in Like-Sign Pairs}

 \bigskip
 \bigskip

 \noindent
{ \bf Claudio Corian\`o\footnote{claudio.coriano@le.infn.it} and Paul H. Frampton\footnote{paul.h.frampton@gmail.com}}

\bigskip

{Dipartimento di Matematica e Fisica "Ennio De Giorgi", \\
Universit\`a del Salento and INFN-Lecce, Via Arnesano, 73100 Lecce, Italy.}

\end{center}

\bigskip
\bigskip

\begin{abstract}
We consider pair production of bileptons $Y^{++}Y^{--}$ at the LHC
for the presently accumulated integrated luminosity of $150/fb$.
It is shown that the entire mass range
800 GeV $\leq$ M(Y) $\leq$ 2000 GeV can be successfully searched.
A bilepton resonance will have an exceptionally large ratio of signal
to background because the standard model prediction is so infinitesimal.
A $5\sigma$ discovery is quite feasible.
\end{abstract}

\end{titlepage}

\renewcommand{\thepage}{\arabic{page}}
\setcounter{page}{1}
\renewcommand{\thefootnote}{\#\arabic{footnote}}

\newpage

\section{Introduction}

\noindent
High energy particle physics is in a crisis because of the lack of new
physics discovered at the LHC since the Higgs boson in 2012. This
impasse impacts also on early universe studies in cosmology and thus
effects astrophysics and astronomy; and progress in particle theory
has traditionally been intertwined with that in condensed matter
physics.

\bigskip

\noindent
Because the LHC data have provided no encouragement for the most popular
extensions beyond the standard model (BSM), we are led to consider less
popular alternatives. One attractive and conservative possibility is to adhere
to the way in which the standard model was discovered, {\it viz} to seek a
renormalizable gauge theory in flat four-dimensional commuting spacetime.

\bigskip

\noindent
One wants the BSM to subsume the standard model and be identical to it at
low energies except for tiny corrections. One also wants it to answer some
mystery about the standard model, {\it e.g.} why there are three quark-lepton
families. Both of these requirements are fulfilled by the Bilepton Model
invented by Frampton\cite{PHF}, and by Pisano and Pleitez\cite{PP}.
There have been two changes in nomenclature since 1992: first, the
original name ``dilepton" was changed\cite{dilepton} in 1996 to ``bilepton" because
of a pre-existing usage for opposite-sign lepton pairs by experimentalists;
second, the original name 331-Model has more recently been changed to Bilepton Model
to mean only that subclass of 331-Models which predict the doubly-charged 
gauge bosons that are the subject of the present article.

\bigskip

\noindent
The reason for the resurgence of interest in the Bilepton Model which
has been cited more in the last decade than previously is probably related
to the lack of encouragement from LHC data for the more popular BSM
models. There has recently been detailed analysis of its properties
\cite{Buras1,Buras2,Buras3,Buras4,Buras5,Buras6}, especially
its predictions with respect to flavour. As stated in \cite{Buras2,Buras3},
one nice feature of these models is the small number of free parameters.

\bigskip

\noindent
In the papers \cite{CCCF1,CCCF2} we have discussed the signatures
of bilepton production at the LHC. At present, the LHC has accumulated
an integrated luminosity of about 150/fb, compared to the expected final
integrated luminosity of 3000/fb. The LHC will be shut down for all of 2019 and
2020 so, for the moment, we must make do with the 150/fb. In \cite{CCCF1},
we studied the pair production of bileptons with two jets detected in the
final state; in \cite{CCCF2}, this was extended to the case of zero jets
detected in the final state. In both cases, the bilepton signal was clearly
visible over the Standard Model background.

\newpage

\section{Resonance bumps}

\bigskip

\noindent
In the present article, we discuss the detection of the bilepton as a
resonance in like-sign lepton pairs. At first sight, with only 150/fb and
the requirement of pair production of the heavy bileptons this might
appear to be an overly optimistic quest. To anticipate our result of
making an approximate, but plausible, estimate of the cross-sections,
the pleasant surprise is that the already-acquired data stored in the
LHC ``cloud" are more than sufficient to
explore the complete range of possible bilepton masses.

\bigskip

\noindent
What is the complete range of bilepton masses? The lower bound comes
from low-energy experiments performed over 20 years ago at PSI who
examined over one trillion muon decays to look for contamination by
right-handed $(V + A)$ interactions of the type which would be generated
by bilepton exchange. The result was summarised by a lower limit on the
Michel parameter and translated into $M(Y) > 800$ GeV. A second
experiment, also at PSI, studied over ten billion examples of muonium
and searched for muonium-antimuonium conversion which can be
mediated by bilepton exchange. By coincidence, this second experiment
provided the same lower limit $M(Y) > 800$ GeV.
The upper limit on $M(Y)$ arises from the fact that, as first pointed out
in \cite{PHF}, there exists an upper limit on the possible scale of
spontaneous symmetry breaking of the 331 gauge group into the 321
gauge group of the standard model. That limit arises from the avoidance
of imaginary couplings which occur if $\sin^2 \theta_{EW} \geq \frac{1}{4}$,
where $\theta_{EW}$ is the electroweak mixing angle. This is an
interesting property of the Bilepton Model because it requires the new
physics to be accessible to the LHC. The maximum 331 breaking scale
is 4000 GeV. One then argues that the dynamics are analagous to
the electroweak model and the Y is analagous to the W. Hence the
expected bilepton mass is $4000 \times (80/248) \simeq 1300$ GeV.
By this argument, it is improbable that $M(Y) > 2000$ GeV.
Thus, our five benchmark points will be the equal spaced masses
$M(Y) = 800, 1100, 1400, 1700, 2000$ GeV.

\bigskip

\noindent
Our method of estimating the numbers of events at ATLAS is to use
old ATLAS data \cite{ATLASzprime} which were analysed to search for a
SSM(=Sequential Standard Model) $Z^{'}$ which was not discovered.
There is some similarity between production of Y and $Z^{'}$. The biggest
difference is that Y must be pair produced so we approximate by using
$\sigma_{SSM}^{Z^{'}} (M(Z^{'} =2M(Y))$. We need to estimate the
brancing ratio (BR) for $(Y \rightarrow e^+e^+, \mu^+\mu^+, \tau^+\tau^+)$.
Because there exist non-leptonic decays $B\rightarrow Q\bar{q}, q\bar{Q}$
where Q is an exotic quark, the BRs depend on the mass M(Q)
of the exotic quarks.

\bigskip

\section{Exotic quark mass}

\noindent
In the Bilepton Model there are three exotic quarks $Q^i$ with i=1,2,3,
one for each familiy. $Q^1$ and $Q^2$ have electric charge $-\frac{4}{3}$
while $Q^3$ has $+\frac{5}{ 3}$. For simplicity, we shall assume all
three have the same mass $M(Q^1)=M(Q^2)=M(Q^3) = M(Q)$.

\bigskip

\noindent
The value of $M(Q)$ is regretfully not known from experiment and it is
important in predicting our event rates. 

\bigskip

\noindent
In what we shall call (A), $M(Q)=0$, and the branching ratios
for $Y$ decay resemble those for a sequential $Z^{'}$, meaning that
$BR = 0.03$. In this limit the S=signal and B=background events
are predicted as shown in Table 1.

\bigskip

\begin{table}[h!]
\caption{Numbers of signal and background events at resonance in like-sign
lepton pairs, calculated as explained in the text (A) for integrated
luminosity 150/fm.}
\begin{center}
\begin{tabular}{||c|c|c|c|c||}
\hline
M(Y) & $\sigma^{Z'}_{SSM}(M_{Z'}=2M(Y))$ & $(BR)^2$ & S=signal & B=background  \\
GeV & (fb)& & events &  events. \\
\hline
\hline
800 & 100  & 0.0009 &13.5 & $< 0.01$  \\
\hline
1100  & 20 & 0.0009 & 2.7 & $<0.01$  \\
\hline
1400 &  6 & 0.0009 & 0.81 & $< 0.01$ \\
\hline
1700 & 1 & 0.0009 & 0.135 & $<0.01$ \\
\hline
2000 & 0.6  & 0.0009 & 0.081 & $<0.01$ \\
\hline
\hline
\end{tabular}
\end{center}
\label{BRs}
\end{table}

\bigskip

\noindent
Here, the predicted S are not encouraging, especially for the highest M(Y). 

\bigskip

\noindent
As another illustration we consider (B) $M(Q) > 2000$ GeV so that the
non-leptonic $Y$ decay is kinematically excluded and $BR = 0.33$ with
the results shown in Table 2.

\begin{table}[h!]
\caption{Numbers of signal and background events at resonance in like-sign
lepton pairs, calculated as explained in the text (B) for integrated
luminosity 150/fm.}
\begin{center}
\begin{tabular}{||c|c|c|c|c||}
\hline
M(Y) & $\sigma^{Z'}_{SSM}(M_{Z'}=2M(Y))$ & $(BR)^2$ & Signal & Background  \\
GeV & (fb)& & events &  events. \\
\hline
\hline
800 & 100  & 0.109 &1635 & $< 0.01$  \\
\hline
1100  & 20 & 0.109 & 327 & $<0.01$  \\
\hline
1400 &  6 & 0.109 & 29 & $< 0.01$ \\
\hline
1700 & 1 & 0.109 & 16 & $<0.01$ \\
\hline
2000 & 0.6  & 0.109 & 6 & $<0.01$ \\
\hline
\hline
\end{tabular}
\end{center}
\label{BRs}
\end{table}

\bigskip

\noindent
Here, the predictions for S are much more hopeful. The question
is what is the case in the real world?

\section{Most reliable predictions}

\noindent
In the previous subsection we investigated two extremes where
(A)$M(Q)=0, BR=0.03$ and (B)$M(Q)=2000$ GeV, BR=0.33 respectively.
These are two limits and the real world lies between (A) and (B). We shall
designate our most reliable estimates by (C) in which $M(Q)=800$ GeV.
In this case, we have $BR=0.33$ when $M(Y)=800$ GeV because the
non-leptonic decay is kinematically disallowed. For the higher values
of $M(Y)$, BR decreases as dictated by 2-body phase space, according
to Table 3.

\bigskip

\noindent

\begin{table}[h!]
\caption{Branching ratio for bilepton $Y^{++}$ decay into like-sign leptons,
with BR($Y^{++} \rightarrow \tau^+\tau^+$)
=BR($Y^{++} \rightarrow \mu^+\mu^+$)=BR($Y^{++} \rightarrow e^+e^+$). ~~~ Assumes
exotic quark mass $M(Q)=800$ GeV for all three families.}
\begin{center}
\begin{tabular}{||c|c||}
\hline
M(Y) & BR=branching ratio   \\
GeV & into like-sign lepton pair \\
   & (per flavour) \\
\hline
\hline
800 & 0.33  \\
\hline
1100  &  0.31 \\
\hline
1400 & 0.28 \\
\hline
1700 & 0.25 \\
\hline
2000 & 0.21 \\
\hline
\hline
\end{tabular}
\end{center}
\label{BRs}
\end{table}

\bigskip

\noindent
In case (C) we re-calculate the numbers of S=signal events and 
B=background events, using the values of BR from Table 3. 
The results which are our most reliable predictions for the LHC
are displayed in Table 4.

\begin{table}[h!]
\caption{Numbers of signal and background events at resonance in like-sign
lepton pairs, calculated as explained in the text (C) for integrated
luminosity 150/fm. This Table gives our most reliable predictions.}
\begin{center}
\begin{tabular}{||c|c|c|c|c||}
\hline
M(Y) & $\sigma^{Z'}_{SSM}(M_{Z'}=2M(Y))$ & $(BR)^2$ & Signal & Background  \\
GeV & (fb)& & events &  events. \\
\hline
\hline
800 & 100  & 0.109 &1635 & $< 0.01$  \\
\hline
1100  & 20 & 0.096 & 288 & $<0.01$  \\
\hline
1400 &  6 & 0.078 & 70 & $< 0.01$ \\
\hline
1700 & 1 & 0.062 & 9 & $<0.01$ \\
\hline
2000 & 0.6  & 0.044 & 4 & $<0.01$ \\
\hline
\hline
\end{tabular}
\end{center}
\label{BRs}
\end{table}

\newpage

\bigskip

\noindent
Because of the insignificant standard model background, for any
of these $Y^{\pm\pm}$ masses, detection of events at 
resonance signals discovery of a
new particle.

\section{Summary}

\noindent
By analysis of the already existing LHC data, corresponding to an integrated
luminosity 150/fb, the doubly-charged bileptonic gauge bosons predicted
by the Bilepton Model can potentially be discovered for the entire mass
range from 800 GeV to 2000 GeV.

\bigskip

\noindent
Our results for our most reliable predictions are summarised in Table 4,
Because background is tiny, a $5\sigma$ signal is
quite feasible.

\bigskip

\noindent
Such a discovery can revolutionise particle physics by providing 
direction for research beyond the fifty-year-old standard
model and giving useful knowledge about the evolution
of the early universe.

\begin{center}
\section*{Acknowledgement}
\end{center}
\noindent
We thank G. Corcella and A. Costantini for useful discusssions. The work
of C.C. is partially supported by INFN
with Iniziativa Specifica QFT-HEP.

\end{document}